\documentclass[10pt,confenrence]{IEEEtran}
\IEEEoverridecommandlockouts

\usepackage{cite}
\usepackage{amsmath,amssymb,amsfonts}
\usepackage{algorithmic}
\usepackage{graphicx}
\usepackage{textcomp}
\usepackage{xcolor}
\usepackage{url}
\usepackage{enumitem}
\usepackage{multirow}
\usepackage{booktabs}
\usepackage[colorlinks,linkcolor=blue]{hyperref}

\def\BibTeX{{\rm B\kern-.05em{\sc i\kern-.025em b}\kern-.08em
    T\kern-.1667em\lower.7ex\hbox{E}\kern-.125emX}}
\begin{document}

\title{3D-Speaker-Toolkit: An Open-Source Toolkit for Multimodal Speaker Verification and Diarization}

\author{\IEEEauthorblockN{
Yafeng Chen$^{1}$, Siqi Zheng$^{1}$, Hui Wang$^{1}$, Luyao Cheng$^{1}$, Tinglong Zhu$^{2}$, Rongjie Huang$^{3}$, \\
Chong Deng$^{1}$, Qian Chen$^{1}$, Shiliang Zhang$^{1}$, Wen Wang$^{1}$, Xihao Li$^{4}$}
\\

\IEEEauthorblockA{\textit{$^{1}$Speech Lab, Alibaba Group}~~~~\textit{$^{2}$Carnegie Mellon University}}
\\
\IEEEauthorblockA{\textit{$^{3}$Zhejiang University}~~~~\textit{$^{4}$University of North Carolina at Chapel Hill}}
}

\maketitle

\begin{abstract}
We introduce \textbf{3D-Speaker-Toolkit}, an open-source toolkit for multimodal speaker verification and diarization, designed for meeting the needs of academic researchers and industrial practitioners. The 3D-Speaker-Toolkit adeptly leverages the combined strengths of acoustic, semantic, and visual data, seamlessly fusing these modalities to offer robust speaker recognition capabilities. The acoustic module extracts speaker embeddings from acoustic features, employing both fully-supervised and self-supervised learning approaches. The semantic module leverages advanced language models to comprehend the substance and context of spoken language, thereby augmenting the system's proficiency in distinguishing speakers through linguistic patterns. The visual module applies image processing technologies to scrutinize facial features, which bolsters the precision of speaker diarization in multi-speaker environments. Collectively, these modules empower the 3D-Speaker-Toolkit to achieve substantially improved accuracy and reliability in speaker-related tasks. With 3D-Speaker-Toolkit, we establish a new benchmark for multimodal speaker analysis. The toolkit also includes a handful of open-source state-of-the-art models and a large-scale dataset containing over 10,000 speakers. The toolkit is publicly available at https://github.com/modelscope/3D-Speaker.
\end{abstract}


\section{Introduction}
\label{sec:intro}
The research of speaker representation learning has achieved remarkable progress in recent years~\cite{DBLP:conf/icassp/SnyderGSPK18,
DBLP:conf/icassp/WanWPL18, DBLP:conf/interspeech/DesplanquesTD20, DBLP:conf/slt/ZhouZW21, DBLP:conf/nips/0004LW023}. Speaker representation learning captures the unique characteristics of a speaker in a compact form, and is extensively utilized across various tasks including speaker verification~\cite{DBLP:conf/interspeech/DesplanquesTD20}, speaker diarization~\cite{DBLP:journals/csl/ParkKDHWN22}, target speaker extraction~\cite{DBLP:journals/spm/ZmolikovaDOKCY23}, speaker-attributed automatic speech recognition~\cite{DBLP:conf/icassp/YuZGFDZHXTWQLYM22}, and other speech-related tasks~\cite{DBLP:conf/icml/CasanovaWSJGP22, DBLP:conf/icassp/Yidi}. Researchers have studied different training schemes to learn robust speaker representations, from fully-supervised leaning~\cite{DBLP:conf/interspeech/DesplanquesTD20,DBLP:conf/icassp/SnyderGSPK18} to self-supervised learning~\cite{DBLP:journals/jstsp/ZhangY22,DBLP:conf/slt/ChenQHQZ22,DBLP:conf/icassp/SangLLAW22}. With the continuous emergence of open-source projects, speech-related technology has advanced in sophistication. Table~\ref{tab:toolkits} lists some well-known open-source speech toolkits. Initially, foundational speech processing toolkits such as Kaldi~\cite{povey2011kaldi} serve as crucial resources for both researchers and industrial applications. More recently, ESPnet~\cite{DBLP:conf/interspeech/WatanabeHKHNUSH18} and SpeechBrain~\cite{ravanelli2021speechbrain} provide novice-friendly code designs and usage. Furthermore, VoxCeleb\_Trainer~\cite{DBLP:conf/interspeech/ChungHMLHCHJLH20}, ASV-Subtools~\cite{DBLP:conf/icassp/TongZZLL0H21}, and Wespeaker~\cite{DBLP:conf/icassp/WangLWCZXDQ23} offer open-source platforms for the speaker community to effortlessly build models.

\begin{table}[t!]
  \caption{Comparison between our 3D-Speaker-Toolkit and well-known open-source toolkits on their support for model deployment and multimodality modeling capabilities. 3D-Speaker-Toolkit is a novel \textbf{multimodal} toolkit that jointly utilizes acoustic, semantic, and visual information to enhance the performance of speaker-related tasks.} 
  \label{tab:toolkits}
  \centering
  \setlength{\tabcolsep}{3.0pt}
  \begin{tabular}{ccc}
    \toprule
    \textbf{Framework} & \textbf{Deployment Support} & \textbf{Modality}\\
    \midrule
    \href{https://github.com/kaldi-asr/kaldi}{Kaldi}\cite{povey2011kaldi} & No & Acoustic \\
    \href{https://github.com/espnet/espnet}{ESPnet}\cite{DBLP:conf/interspeech/WatanabeHKHNUSH18} & Yes & Acoustic \\
    \href{https://github.com/speechbrain/speechbrain}{SpeechBrain}\cite{ravanelli2021speechbrain} & No & Acoustic \\
    \href{https://github.com/clovaai/voxceleb_trainer}  {VoxCeleb\_Trainer}\cite{DBLP:conf/interspeech/ChungHMLHCHJLH20} & No & Acoustic \\
    \href{https://github.com/Snowdar/asv-subtools}{ASV-Subtools}\cite{DBLP:conf/icassp/TongZZLL0H21} & No & Acoustic \\
    \href{https://github.com/wenet-e2e/wespeaker}{Wespeaker}\cite{DBLP:conf/icassp/WangLWCZXDQ23} & Yes & Acoustic \\
    \toprule
    \href{https://github.com/alibaba-damo-academy/3D-Speaker/tree/main}{\textbf{3D-Speaker-Toolkit (Ours)}} & Yes & Acoustic+Visual+Textual \\
    \bottomrule
  \end{tabular}
\end{table}

Table~\ref{tab:toolkits} compares 3D-Speaker-Toolkit to well-known speech toolkits. To the best of our knowledge, 3D-Speaker-Toolkit
is the first open-source speaker toolkit that extends beyond the acoustic dimension and pioneers a comprehensive approach by integrating acoustic, semantic, and visual modalities for a multifaceted analysis of speaker identity and characteristics. 3D-Speaker-Toolkit releases code and models that leverage multimodal information to improve accuracy and reliability for speaker-related tasks. The aim of the toolkit is to provide researchers with a robust and flexible platform for developing, training, and deploying state-of-the-art (SOTA) models, thereby accelerating research and deployment. The highlights of the 3D-Speaker-Toolkit are summarized as follows:

\begin{itemize}[leftmargin=*]
    \item \textbf{Multimodality}: 
    Current mainstream speaker-related approaches primarily rely on acoustic information; however, they may suffer from degraded performance under adverse acoustic conditions. To address this limitation, we propose multimodal techniques that can effectively fuse information from audio, video, and text modalities~\cite{DBLP:conf/acl/ChengZZWCC23, DBLP:journals/corr/abs-2309-10456,cheng2024integrating} for speaker-related tasks. Our proposed method achieves substantial improvement over conventional acoustic-only approaches.

    \item \textbf{Off-the-shelf Usage and Production Ready}: 3D-Speaker-Toolkit provides dozens of speaker embedding extractors on ModelScope\footnote{\url{https://modelscope.cn/models}}. These extractors are trained on public and large-scale in-house data, and achieve performance that suffices production usage. As to \textit{Deployment Support} in Table~\ref{tab:toolkits}, all models in 3D-Speaker-Toolkit can be exported in the ONNX format for straightforward adoption in deployment environments, similar to ESPnet and Wespeaker.

    \item \textbf{Large-Scale Training Data}: In 3D-Speaker-Toolkit, we also release a large-scale speech corpus, \textbf{3D-Speaker dataset}~\cite{DBLP:journals/corr/abs-2306-15354}.  This dataset contains over 10,000 speakers, partitioned into training and test sets.  Each speaker is simultaneously recorded by multiple devices, with the distance from the speakers to the recording devices varying to cover most common scenarios. Some speakers also speak multiple dialects. The controlled combinations of this multi-dimensional audio data yield a matrix of diverse blends of speech representation entanglement, thereby motivating effective methods to untangle them.
    
    \item \textbf{State-of-the-art Performance}: We release a set of training and inference recipes in 3D-Speaker-Toolkit for state-of-the-art models~\cite{ DBLP:conf/interspeech/WangZCC023, DBLP:conf/interspeech/ChenZWC0Q23, DBLP:conf/interspeech/DesplanquesTD20,DBLP:journals/pami/GaoCZZYT21, DBLP:conf/cvpr/HeZRS16, chen2024eres2netv2} for both speaker verification and diarization tasks. We achieve competitive performance on several popular benchmarks, including VoxCeleb~\cite{DBLP:journals/csl/NagraniCXZ20}, CN-Celeb~\cite{DBLP:conf/icassp/FanKLLCCZZCW20, DBLP:journals/speech/LiLKFCCVZW22}, and our 3D-Speaker dataset~\cite{DBLP:journals/corr/abs-2306-15354}.

    \item \textbf{SSL Support}: Training speaker-discriminative and robust speaker verification systems without explicit speaker labels remains a persistent challenge. To tackle this challenge, 3D-Speaker-Toolkit includes implementations of several competitive self-supervised learning (SSL) algorithms that greatly reduce reliance on labeled data, including DINO~\cite{DBLP:conf/iccv/CaronTMJMBJ21}, RDINO~\cite{DBLP:conf/icassp/ChenZWCC23}, and SDPN~\cite{DBLP:journals/corr/abs-2308-02774}.

    \item \textbf{Lightweight}: Designed for speaker-related tasks, 3D-Speaker-Toolkit features clean and simple code that is entirely based on PyTorch
    and its ecosystem. Consequently, independent of Kaldi~\cite{povey2011kaldi}, 3D-Speaker-Toolkit simplifies installation and usage and provides lightweight solutions.

\end{itemize}

\section{3D-Speaker-Toolkit}
\label{sec:3dspeaker}

\begin{figure}[t]
  \centering
  \includegraphics[scale=0.55]{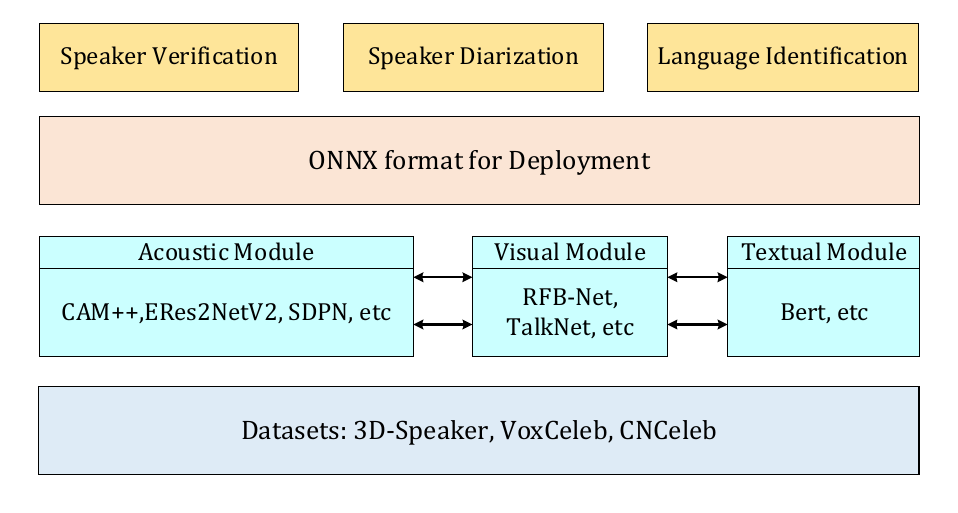}
  \caption{The overall design of the 3D-Speaker-Toolkit. It utilizes acoustic, visual, and textual information for speakers, supports both supervised and self-supervised models, includes recipes for model export and deployment, and facilitates multiple speaker-related tasks.}
  \label{fig: overall}
\end{figure}

3D-Speaker-Toolkit is committed to open and transparent science. It primarily comprises three components as \textit{acoustic}, \textit{semantic}, and \textit{visual} modules. We train models using publicly available datasets to ensure that our results can be easily replicated. To facilitate off-the-shelf usage, we offer a wide range of pre-trained models that are readily accessible and can be utilized with just a few lines of code. The overall design of 3D-Speaker-Toolkit is illustrated in Fig.~\ref{fig: overall}.

\subsection{Acoustic Module}
This module introduces traditional acoustic information to extract discriminative speaker representation which can be used in speaker verification, diarization and so on. We detail the feature preparation, model training, model export and deployment, and model inference as follows.

\subsubsection{Feature Preparation}
Traditional feature preparation for speaker embedding learning is commonly performed offline. These methods require static training samples that are stored on disk and do not change during the training stage. In contrast, 3D-Speaker-Toolkit processes the original waveform data in real time, which offers two main advantages: Firstly, the real-time online processing eliminates the need to store augmented waveforms and processed features, hence significantly reduces disk usage requirements. Secondly, real-time augmentation exposes the model to varied training samples in each epoch, which could inject variability and randomness into training and hence improve the robustness of the models.

\subsubsection{Supported Models}

3D-Speaker-Toolkit supports a variety of models for speaker-related tasks, including fully supervised models such as ECAPA-TDNN~\cite{DBLP:conf/interspeech/DesplanquesTD20}, ResNet34~\cite{DBLP:conf/cvpr/HeZRS16}, Res2Net~\cite{DBLP:journals/pami/GaoCZZYT21}, ERes2Net~\cite{DBLP:conf/interspeech/ChenZWC0Q23}, ERes2NetV2\cite{chen2024eres2netv2}, CAM++~\cite{DBLP:conf/interspeech/WangZCC023}, and self-supervised models such as DINO\cite{DBLP:conf/iccv/CaronTMJMBJ21}, RDINO~\cite{DBLP:conf/icassp/ChenZWCC23}, and SDPN~\cite{DBLP:journals/corr/abs-2308-02774}. Classic models such as ResNet34, Res2Net, ECAPA-TDNN, and DINO are well-established and will not be elaborated here. Notably, our own models, namely \textbf{ERes2Net}, \textbf{ERes2NetV2}, \textbf{CAM++}, \textbf{RDINO}, and \textbf{SDPN}, bring innovative approaches to the field. We plan to add more SOTA models into 3D-Speaker-Toolkit.

Among our innovations in fully-supervised models, ERes2Net enhances performance through fusion of both local and global features on the basis of Res2Net. Its successor, ERes2NetV2, is tailored to more effectively capture features from short-duration utterances. CAM++ is built on a densely connected Time Delay Neural Network (D-TDNN)~\cite{DBLP:conf/interspeech/YuL20} backbone and employs a novel multi-granularity pooling technique~\cite{DBLP:conf/iclr/TanCWZZL22} to capture contextual information at various levels with reduced computational complexity.

Regarding our innovative self-supervised models, RDINO introduces two regularization terms applied to embeddings within DINO to mitigate the model collapse problem in non-contrastive self-supervised speaker verification frameworks. SDPN framework assigns representations of augmented views of utterances to the same prototypes as the original view, facilitating learning speaker-discriminative self-supervised speaker representations.

\subsubsection{Model Export and Deployment}
For models trained with 3D-Speaker-Toolkit, exporting them in the ONNX format for deployment on the Triton Inference Server is straightforward. Additionally, we offer off-the-shelf usage for the models released in the toolkit.
Users can easily load a pre-trained speaker embedding extractor by specifying the model's name.

\subsubsection{Embedding Extraction and Inference}
With deployed feature extractors, users can quickly extract speaker embeddings with just a few lines of code. Additional processing is applied at the score level. After deriving all scores for trials, score normalization is applied according to the configuration. 

\subsection{Multimodal Module}
In real-world scenarios, the performance of audio-only systems often suffers from low-quality acoustic environments, which are typically characterized by the presence of background noise, reverberation, and overlapping speech from multiple speakers.
It is known that visual and semantic cues, such as facial activities and dialogue patterns, can enhance human perception of auditory information, helping identify the current active speakers.
Recently, much speaker-related research has focused on integrating visual or semantic information into acoustic-only systems, achieving performance improvement~\cite{DBLP:journals/csl/ParkKDHWN22, e2e-av-sd, who-said-that,Kanda2021TranscribetoDiarizeNS}.
Despite the rapid advancements in multimodal modeling, numerous challenges remain, such as unreliable visual information, occluded or off-screen speakers, and the complexity of natural conversation scenarios~\cite{chung2020spot, xu2022ava}. Currently, 3D-Speaker-Toolkit is primarily focused on the multimodal speaker diarization task. Looking ahead, we will extend the toolkit to the multimodal speaker verification task. The textual and visual information leveraged in multimodal speaker diarization can be utilized similarly in multimodal speaker verification.

\begin{figure}[t]
  \centering
  \includegraphics[scale=0.25]{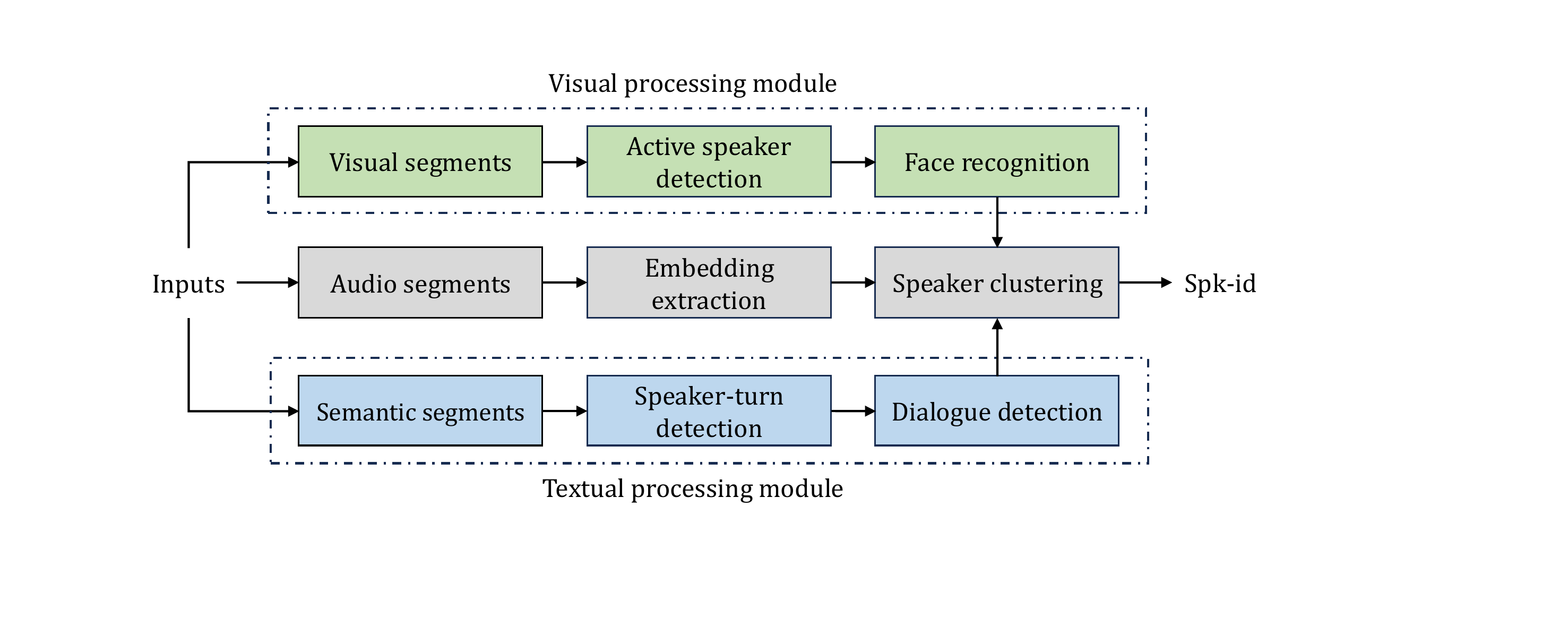}
  \caption{Overview of multimodal speaker diarization system. An overview of multimodal speaker diarization system. It incorporates additional visual and textual processing modules that independently extract visual and semantic information to provide guidance for the audio-only diarization.}
  \label{fig:audio-visual-SD}
\end{figure}

Speaker diarization is the process to identify ``who spoke when" in a given multi-person conversation. Traditional diarization technology consists of voice activity detection, speech segmentation, embedding extraction, and clustering. However, this process may be compromised by a complex acoustic environment. We present a modular multimodal speaker diarization system that uses independent module for each type of observation and fuses this information within a unified clustering framework.
The overview of the system is shown in Fig.~\ref{fig:audio-visual-SD}. 

\subsubsection{Visual Module}
Alongside the traditional acoustic-only diarization process, we incorporate a visual diarization pipeline including face tracking, active speaker detection and face recognition, similar to~\cite{chung2020spot, xu2022ava}. 
The face tracking module uses models such as RFB-Net~\cite{DBLP:conf/eccv/LiuHW18} to detect and locate faces within the video content over time, thereby creating a consistent track for each face present. Only those face tracks that correspond with audible speech segments, as identified by VAD, are retained for further processing.
Then active speaker detection such as TalkNet~\cite{DBLP:conf/mm/TaoPDQS021} takes the cropped face video and corresponding audio as input and decides whether the tracked faces correspond to active speakers at any given moment. In order to ensure the effectiveness of the subsequent process, frames of low quality are then filtered out by employing a face quality assessment model.
Subsequently, a face recognition model, such as CurricularFace~\cite{DBLP:conf/cvpr/HuangWT0SLLH20}, is utilized to extract embeddings for every face track. These embeddings are extracted at uniform intervals from each face track. These are then clustered using Agglomerative Hierarchical Clustering (AHC).
Finally, each face, along with its clustering label, can be aligned with respective acoustic embeddings along the time axis and used to provide guidance for audio-only clustering. 

\begin{table*}[t!]
    \caption{Results on VoxCeleb and 3D-Speaker datasets for fully-supervised speaker verification. We use the development set of VoxCeleb2 and the training set of 3D-Speaker for training, respectively. The 3D-Speaker test sets include three trials: Multi-Device, Multi-Distance, and Multi-Dialect. The best results are boldfaced.}
    \vspace{-3mm}
    \label{tab:data_augmentation}
    \centering
    \setlength\tabcolsep{2.6pt}
    \begin{tabular}{c c c c c c c c c c c c c}
    \toprule
    \multirow{2}{*}{} & \multicolumn{6}{c}{\textbf{VoxCeleb}} & \multicolumn{6}{c}{\textbf{3D-Speaker}} \\
    \cmidrule(lr){2-7} \cmidrule(lr){8-13}
    \multirow{2}{*}{} & \multicolumn{2}{c}{\textbf{VoxCeleb1-O}} & \multicolumn{2}{c}{\textbf{VoxCeleb1-E}} & \multicolumn{2}{c}{\textbf{VoxCeleb1-H}} & \multicolumn{2}{c}{\textbf{Multi-Device}} & \multicolumn{2}{c}{\textbf{Multi-Distance}} & \multicolumn{2}{c}{\textbf{Multi-Dialect}} \\
    \cmidrule(lr){2-3} \cmidrule(lr){4-5} \cmidrule(lr){6-7} \cmidrule(lr){8-9} \cmidrule(lr){10-11} \cmidrule(lr){12-13} 
    & \textbf{EER(\%)}$\downarrow$ & \textbf{MinDCF}$\downarrow$ & \textbf{EER(\%)}$\downarrow$ & \textbf{MinDCF}$\downarrow$ & \textbf{EER(\%)}$\downarrow$ & \textbf{MinDCF}$\downarrow$ & \textbf{EER(\%)}$\downarrow$ & \textbf{MinDCF}$\downarrow$ & \textbf{EER(\%)}$\downarrow$ & \textbf{MinDCF}$\downarrow$ & \textbf{EER(\%)}$\downarrow$ & \textbf{MinDCF}$\downarrow$ \\
    \midrule
    Res2Net & 1.56 & 0.150 & 1.41 & 0.149 & 2.48 & 0.230 & 8.03 & 0.707 & 9.67 & 0.781 & 14.11 & 0.920 \\
    ResNet34 & 1.05 & 0.107 & 1.11 & 0.116 & 1.99 & 0.192 & 7.29 & 0.689 & 8.98 & 0.762 & 12.81 & 0.906 \\
    ECAPA-TDNN & 0.86 & 0.116 & 0.97 & 0.112 & 1.90 & 0.193 & 8.55 & 0.728 & 12.15 & 0.814 & 12.24 & 0.915 \\
    ERes2Net & 0.84 & 0.088 & 0.96 & 0.102 & 1.78 & 0.175 & 7.12 & 0.657 & 9.82 & 0.749 & 12.10 & 0.866 \\
    CAM++ & 0.65 & 0.086 & 0.81 & 0.094 & 1.58 & 0.163 & 7.17 & 0.669 & 9.84 & 0.722 & 11.78 & 0.844 \\
    ERes2NetV2 & \textbf{0.61} & \textbf{0.054} & \textbf{0.76} & \textbf{0.082} & \textbf{1.45} & \textbf{0.143} & \textbf{6.52} & \textbf{0.589} & \textbf{8.88} & \textbf{0.690} & \textbf{11.30} & \textbf{0.825} \\
    \bottomrule
    \end{tabular}
\end{table*}

\begin{table}[th]
    \caption{Results on CN-Celeb dataset. The development sets of CN-Celeb1 and CN-Celeb2 are used for training. We compare the number of parameters (Params) and floating-point operations (FLOPs) of different models.}
    \vspace{-3mm}
    \label{tab: fully-supervised-cn}
    \centering
    \begin{tabular}{c c c c c }
    \toprule
    \textbf{Framework} & \textbf{Params} & \textbf{FLOPs} & \textbf{EER(\%)}$\downarrow$ & \textbf{MinDCF}$\downarrow$ \\
    \midrule
    Res2Net & 4.03M & 6.85G & 7.69 & 0.452 \\
    ResNet34 & 6.34M & 2.65G & 6.92 & 0.421 \\
    ECAPA-TDNN & 20.7M & 5.64G & 7.67 & 0.442 \\
    CAM++ & 7.18M & 1.72G & 6.30 & 0.370 \\
    ERes2Net & 6.61M & 5.16G & 6.11 & 0.371 \\
    ERes2NetV2 & 17.8M & 12.6G & 6.04 & 0.362 \\
    \bottomrule
    \end{tabular}
\end{table}

\begin{table}[thb]
    \caption{Comparison between self-supervised learning models on VoxCeleb1-O. ``*" denotes the SSL frameworks in 3D-Speaker-Toolkit.}
    \vspace{-3mm}
    \label{tab:compare_ssl}
    \centering
    \setlength{\tabcolsep}{9pt}
    \begin{tabular}{ccc}
    \toprule
    \textbf{Model} & \textbf{Embedding Extractor} & \textbf{EER(\%)}$\downarrow$ \\
    \midrule
    SSReg~\cite{DBLP:conf/icassp/SangLLAW22} & Fast ResNet34 & 6.99 \\
    Mixup-Aug~\cite{DBLP:conf/icassp/ZhangJCLHS22} & Fast ResNet34 & 5.84 \\
    RDINO*~\cite{DBLP:conf/icassp/ChenZWCC23} & ECAPA-TDNN & 3.24 \\
    DINO*~\cite{DBLP:journals/corr/abs-2308-02774} & ECAPA-TDNN & 2.65 \\
    DINO-Aug~\cite{DBLP:conf/slt/ChenQHQZ22} & ECAPA-TDNN & 2.51 \\
    C3-DINO~\cite{DBLP:journals/jstsp/ZhangY22} & ECAPA-TDNN & 2.50 \\
    SDPN* \cite{DBLP:journals/corr/abs-2308-02774} & ECAPA-TDNN & 1.80 \\
    \bottomrule
    \end{tabular}
\end{table}

\subsubsection{Textual Module}
Textual data provide rich contextual and semantic content, which can reveal clear linguistic patterns and identify speaker-turns based on semantic breaks~\cite{DBLP:conf/acl/ChengZZWCC23}. To facilitate extracting speaker information from text, we design two semantic tasks: dialogue detection and speaker-turn detection. Dialogue detection is formulated as a binary classification task to determine whether a text segment is a dialogue or not. On the other hand, speaker-turn detection is a sequence labeling task aimed at identifying the locations of speaker change in the text. Our current design consists of calling a dialogue detection system first to ascertain whether a segment of text constitutes a conversation. For those conversation segments, a speaker-turn detection module is further applied to pinpoint the location where a speaker change occurs. Currently, these models are trained on the AISHELL-4~\cite{DBLP:journals/corr/abs-2104-03603} and AliMeeting~\cite{DBLP:conf/icassp/YuZGFDZHXTWQLYM22} datasets based on BERT~\cite{DBLP:conf/naacl/DevlinCLT19}. Results from this pipeline enable encapsulation of speaker-related textual information. Our experiments show that the pipelined approach with these two models outperforms solely using speaker-turn detection module in the overall multimodal speaker diarization. Within 3D-Speaker-Toolkit, we open-source these two models for extracting speaker-related semantic information. Notably, the extensible and modularized design of our toolkit supports easy replacement of the textual module with new modules based on different frameworks and using advanced language models, including large language models, as the backbone.

\subsubsection{Audio-visual-textual Module}

The inherent limitations of each individual modality constrain the efficacy of \textit{unimodal} speaker diarization. On the other hand, each modality offers distinct and complementary strengths.
Therefore,  we focus on developing a unified framework that simultaneously leverages audio, visual, and semantic cues. Specifically, we employ a clustering method based on constrained optimization. By carefully constructing visual and semantic constraints, multimodal information can be effectively integrated through the process of joint constraint propagation using the E2CP method~\cite{Lu2011ExhaustiveAE}. For more details, please refer to~\cite{cheng2024integrating}.

\section{Experiments}
\label{sec:experiments}
\subsection{Speaker verification}
For the speaker verification task, we construct recipes based on the VoxCeleb~\cite{DBLP:journals/csl/NagraniCXZ20}, 3D-Speaker dataset~\cite{DBLP:journals/corr/abs-2306-15354}, and CN-Celeb~\cite{DBLP:conf/icassp/FanKLLCCZZCW20, DBLP:journals/speech/LiLKFCCVZW22} datasets, using two metrics: equal error rate (EER) and the minimum of the normalized detection cost function (MinDCF). The performance of fully-supervised models on the VoxCeleb, 3D-Speaker dataset, and CN-Celeb dataset is listed in Table~\ref{tab:data_augmentation} and Table~\ref{tab: fully-supervised-cn}. 
ERes2Net, CAM++, and ERes2NetV2 results presented here are obtained after large-margin fine-tuning~\cite{DBLP:conf/icassp/ThienpondtDD21}, with cosine scoring. 
Among the six currently supported models, ERes2NetV2 achieves the best performance, while CAM++ yields competitive results with lower computational complexity. The speaker verification performance on the 3D-Speaker dataset is worse than that on the VoxCeleb dataset, primarily due to the fact that 3D-Speaker dataset includes test speech recorded at varying distances, utilizing multiple devices, and encompassing diverse dialects, which pose greater challenges for accurate speaker recognition. 
Table~\ref{tab: fully-supervised-cn} presents the experimental results of models on the CN-Celeb dataset and shows the number of parameters (Params) and floating-point operations (FLOPs) of each model. FLOPs are measured on 3-second long segments. Similar to the trend observed in Table~\ref{tab:data_augmentation}, ERes2NetV2 outperforms the other five models, while CAM++ delivers robust and competitive performance with the additional benefit of lower computational overhead.

The performances of self-supervised models, namely DINO\cite{DBLP:journals/corr/abs-2308-02774}, RDINO\cite{DBLP:conf/icassp/ChenZWCC23}, and SDPN\cite{DBLP:journals/corr/abs-2308-02774}, are reported in Table~\ref{tab:compare_ssl}.
We make a comparison with recently published non-contrastive SSL methods, which includes \cite{DBLP:conf/icassp/SangLLAW22, DBLP:conf/icassp/ZhangJCLHS22, DBLP:conf/slt/ChenQHQZ22}, and the SSL SOTA C3-DINO~\cite{DBLP:journals/jstsp/ZhangY22} that combines contrastive and non-contrastive strategies. On the VoxCeleb1-O test set, our non-contrastive SDPN framework realizes \textbf{1.80\%} EER using the identical cosine distance scoring technique as C3-DINO, surpassing  C3-DINO (2.50\% EER) by \textbf{28.0\%} relative. 

\begin{table}[t]
    \caption{Performance comparison between unimodal and multimodal speaker diarization. The best results are boldfaced.}
    \vspace{-3mm}
    \label{tab: multimodal-SD}
    \centering
    \setlength{\tabcolsep}{3.0pt}
    \begin{tabular}{c c c c c }
    \toprule
    \textbf{Methods} & \textbf{Modality} & \textbf{DER(\%)}$\downarrow$ & \textbf{JER(\%)}$\downarrow$ & \textbf{cpWER(\%)}$\downarrow$ \\
    \midrule
    VBx & Audio & 10.31     & 29.28     & 18.03 \\
    SC & Audio & 9.37   & 27.21         & 17.04 \\
    SC+E2CP & Audio+Visual & 9.13         & 26.02      & 16.83 \\
    SC+E2CP & Audio+Textual & 9.12        & 25.98        & 16.86  \\
    SC+E2CP  & Audio+Visual+Textual & \textbf{9.01} & \textbf{22.57} & \textbf{16.36} \\
    \bottomrule
    \end{tabular}
\end{table}

\subsection{Multimodal Speaker diarization}
Experiments based on the proposed multimodal diarization method are conducted on a self-collected video dataset, which includes 2 to 10 speakers. Strong baselines for audio-only diarization have been established using VBx~\cite{landini2022bayesian} and Spectral clustering(SC)~\cite{von2007tutorial} methods. We use E2CP integrated with spectral clustering method to leverage multimodal information to enhance speaker diarization.
The comparison results are shown in Table~\ref{tab: multimodal-SD}. Common speaker diarization metrics, including Diarization Error Rate(DER)~\cite{Fiscus2006DER}, Jaccard Error Rate(JER)~\cite{Ryant2019TheSD} and concatenated minimum-permutation Word Error Rate(cpWER)~\cite{Watanabe2020CHiME6CT}, are used. The results show that, compared to systems based solely on audio, incorporating visual or textual information both achieve notable performance improvements. Combining all three modalities of audio, visual, and textual data yields superior results on all metrics. 

\section{Conclusion}
In this paper, we introduce the 3D-Speaker-Toolkit, an open-source toolkit that leverages multimodal speaker information to support a range of speaker-related tasks. It is well-designed, lightweight, and demonstrates superior performance on both public and large-scale in-house datasets. Additionally, the 3D-Speaker-Toolkit provides CPU- and GPU-compatible deployment and runtime code. As we move forward, our priorities include effectively adapting large pre-trained models, compressing model sizes, and broadening integration with various speaker-related tasks.

\end{document}